\documentstyle [12pt] {article}
\begin{document}

\pagestyle{empty}
\rightline{\vbox{
\halign{&#\hfil\cr
&NUHEP-TH-94-18 \cr
&August 1994 \cr
&hep-ph/9408273 \cr}}}
\bigskip
\bigskip
\bigskip
{\Large\bf
	\centerline{Next-to-leading Order Debye Mass}
	\centerline{for the Quark-gluon Plasma}
\bigskip
\normalsize

\centerline{Eric Braaten and Agustin Nieto}
\centerline{\sl Department of Physics and Astronomy, Northwestern University,
    Evanston, IL 60208}
\bigskip

\begin{abstract}
The Debye screening mass for a quark-gluon plasma at high temperature
is calculated to next-to-leading order in the QCD coupling constant
from the correlator of two Polyakov loops.  The result agrees with the
screening mass defined by the location of the pole in the gluon propagator
as calculated by Rebhan.  It is logarithmically sensitive to
nonperturbative effects associated with the screening of static
chromomagnetic fields.
\end{abstract}

\vfill\eject\pagestyle{plain}\setcounter{page}{1}

One of the fundamental properties of a plasma is the Debye screening mass
$m_D$,
whose inverse is the screening length for electric fields in the plasma.
The definition of the Debye mass is conventionally given
in terms of the small-momentum ($k \to 0$) limit of the static ($\omega = 0$)
Coulomb propagator $[k^2 + \Pi_{00}(0,k)]^{-1} = [k^2 \epsilon(0,k)]^{-1}$,
where $\Pi_{00}(\omega,k)$ is the longitudinal photon self-energy function
and $\epsilon(\omega,k)$ is the corresponding dielectric function:
\begin{equation}
m_D^2 = \lim_{k\to 0} \Pi_{00}(0,k) .
\label{convdef}
\end{equation}
The Debye mass can alternatively be defined in terms of the location
of the pole in the static propagator for complex $k$:
\begin{equation}
k^2 \;+\; \Pi_{00}(0,k) \;=\; 0 \quad {\rm at} \; k^2 = - m_D^2.
\label{poledef}
\end{equation}
At leading order in the coupling constant, the static self-energy has the
simple form $\Pi_{00}(0,k) = m_D^2$ independent of $k$, and the definitions
(\ref{convdef}) and (\ref{poledef}) are equivalent.  Beyond leading order
in the coupling constant, they need no longer be equivalent.
A fundamental question of plasma physics is then this:
what is the correct general definition of the Debye mass?

This question applies equally well to a quark-gluon plasma, where the
Debye mass describes the screening of chromoelectric fields.
Because the coupling constant of quantum chromodynamics (QCD) is relatively
large, higher order corrections to the Debye mass are probably not negligible.
The question of the correct definition therefore becomes one of practical
importance.  In the case of QCD, the longitudinal gluon self-energy function
$\Pi_{00}(\omega,k)$ is gauge dependent.  If the Debye mass is relevant
to the screening of chromoelectric fields, then it must be gauge invariant.
Thus, gauge invariance can be used as a guide to the correct definition
of the Debye mass.  Formal arguments due to Kobes, Kunstatter, and Rebhan
indicate that, in spite of the gauge-dependence of the gluon propagator,
the locations of its poles are gauge invariant \cite{kkr}.  This
suggests that (\ref{poledef}) is the correct general definition.
However, it is important to back up these formal arguments with explicit
calculations.

There have been a number of calculations of the next-to-leading order
correction to the Debye mass using the conventional definition (\ref{convdef})
\cite{kk,nadkarni}.  The results are infrared finite but
gauge dependent.  The naive application of the pole definition
(\ref{poledef}) gives a gauge-dependent result,
which is also infrared-divergent.
However, it was recently shown by Rebhan \cite{rebhan} that, with a careful
treatment of infrared effects, the pole definition does in fact give a
gauge-invariant result.

Since the calculation of Rebhan involves subtle
interchanges of limits, it is desirable to have an independent
verification of this result.  It is also desirable to calculate
the next-to-leading order Debye mass directly from
a gauge-invariant quantity, to provide assurance that it is relevant to
physical quantities.  The simplest such quantity is the correlation function
of two Polyakov loops.  A previous calculation of this correlator to
next-to-leading nontrivial order by Nadkarni \cite{nadkarni} gave results
that seemed to be incompatible with simple Debye screening.
In this Letter, we reexamine this calculation and show that, by careful
treatment of infrared effects, it can be used to extract the Debye mass to
next-to-leading order.  The result agrees with that obtained by Rebhan
from the pole in the gluon propagator.  These results, together with
the general arguments of Ref. \cite{kkr}, provide compelling evidence that
the pole definition (\ref{poledef}) is the correct definition
of the Debye mass beyond leading order in the coupling constant.

We begin by reviewing the calculation by Rebhan.
At leading order in the QCD coupling constant $g$, the Debye mass is
given by a straightforward perturbative calculation:
$m_D = \sqrt{(2 N_c + N_f)/6} \; g T$,
where $N_c=3$, $N_f$ is the number of flavors of light quarks,
and $T$ is the temperature.  At next-to-leading order, it is necessary
to resum perturbation theory by including the Debye mass in the Coulomb
propagator.  The contribution of order $g^3$ to the static longitudinal
gluon self-energy is
\begin{equation}
\delta \Pi_{00}(0,k) \;=\; N_c g^2 T \int {d^3p \over (2 \pi)^3}
\left[ {1 \over {\bf p}^2 + m_D^2}
	+ {2 (m_D^2 - k^2) \over {\bf p}^2 ({\bf q}^2 + m_D^2)}
	- \xi {(k^2 + m_D^2)({\bf p}^2 + 2 {\bf p}\cdot {\bf k})
			\over ({\bf p}^2)^2 ({\bf q}^2 + m_D^2)} \right]
\label{Pi-int}
\end{equation}
where ${\bf q} = {\bf p} + {\bf k}$.  In (\ref{Pi-int}), ultraviolet
divergences are understood to be removed using dimensional regularization.
Integrating over ${\bf p}$, we obtain the result
\begin{equation}
\delta \Pi_{00}(0,k) \;=\; {N_c g^2 T \over 2 \pi} m_D
\left[ {m_D^2 - k^2 \over 2 i m_D k} \log{m_D + i k \over m_D - i k}
	- {\xi + 1 \over 2} \right].
\end{equation}
Upon taking the limit $k \to i m_D$, we find a correction that is
logarithmically divergent and gauge dependent.  However, as pointed
out by Rebhan, for $k$ near $i m_D$, the integral in (\ref{Pi-int})
is extremely sensitive to the infrared region ${\bf p \to 0}$.
He regularized the infrared region by replacing $1/{\bf p}^2$
by $1/({\bf p}^2 + m_{\rm mag}^2)$.  Taking  the limit $k \to i m_D$
in the presence of the regulator and then taking the limit
$m_{\rm mag} \to 0$, he obtained the gauge-invariant result
\begin{equation}
\delta m_D \;=\; {N_c g^2 T \over 4 \pi}
\left[ \log {2 m_D \over m_{\rm mag}} - {1 \over 2} \right].
\label{delmD}
\end{equation}
The logarithm indicates that the Debye mass at next-to-leading order
is weakly sensitive to the screening of the static chromomagnetic
interaction, which is dominated by nonperturbative effects.
The particular regulator used by Rebhan corresponds to a model for these
nonperturbative effects in which they simply shift the pole
in the transverse gluon propagator from $p = 0$ to the imaginary
values $p = \pm i m_{\rm mag}$.
One could equally well use dimensional regularization as the
infrared regulator.  Again the pole definition (\ref{poledef})
gives a gauge-invariant result which diverges as the regulator is removed.

We next review Nadkarni's calculation of the next-to-leading order
correction to the correlator of Polyakov loops.  The calculation
was carried out in static gauge, where the Polyakov loop reduces
to a simple exponential: $\Omega({\bf x}) = \exp(-i g A_4({\bf x})/T)$.
The correlator of two such operators separated by a distance $R$ is
$C_{PL}(R) = \langle {\rm Tr} \Omega(0) {\rm Tr} \Omega({\bf R})
\rangle/N_c^2$.
Nadkarni calculated this correlator to order $g^5$ and found
\begin{equation}
C_{PL}(R) \;=\; 1
\;+\; {(N_c^2 - 1) g^4 \over 8 N_c^2 T^2} V(R)^2
\left[ 1 \;+\; {N_c g^2 T \over m_D} (f_1(R) + f_2(R)) \right]
\label{CPL}
\end{equation}
where $V(R) = e^{- m_D R}/(4 \pi R)$ is the Debye-screened Coulomb potential.
The complete expressions for the functions $f_1(R)$ and $f_2(R)$
are given in Ref. \cite{nadkarni}.
As in the Appendix of Ref.~\cite{nadkarni}, the integrals can be
simplified by expanding out the numerators and cancelling propagator
factors wherever possible.
Keeping only those terms that grow at least linearly with $R$, the
functions reduce to
\begin{eqnarray}
f_1(R)
&=& - 2 m_D V(R)^{-1} \int {d^3k \over (2 \pi)^3}
	e^{-i {\bf k} \cdot {\bf R}} {1 \over (k^2 + m_D^2)^2}
\nonumber \\
&& \quad \quad \quad \quad \quad \quad
\times \int {d^3p \over (2 \pi)^3}  \left( {1 \over {\bf p}^2 + m_D^2}
	\;+\; {4 m_D^2 \over {\bf p}^2 ({\bf q}^2 + m_D^2)} \right)
\label{f1def} \\
f_2(R)
&=& - 2 m_D V(R)^{-2} \int {d^3k \over (2 \pi)^3}
	e^{-i {\bf k} \cdot {\bf R}} \left( k^2 + 2 m_D^2 \right)
\nonumber \\
&& \quad \quad
\times \int {d^3p_1 \over (2 \pi)^3}  \int {d^3p_2 \over (2 \pi)^3}
{1 \over ({\bf p}_1^2 + m_D^2) ({\bf q}_1^2 + m_D^2)
		({\bf p}_2^2 + m_D^2) ({\bf q}_2^2 + m_D^2)
		({\bf p}_1 - {\bf p}_2)^2 }
\label{f2def}
\end{eqnarray}
where ${\bf q}_i = {\bf p}_i + {\bf k}$.  Ultraviolet divergences are
understood to be regularized using dimensional regularization.
Evaluating the integrals in the limit $R \to \infty$, Nadkarni obtained
\begin{eqnarray}
f_1(R) & \longrightarrow & {m_D R \over 4 \pi}
	\left[ - 2 \log(2 m_D R) + 3 - 2 \gamma \right],
\label{f1nad}
\\
f_2(R) & \longrightarrow & {m_D R \over 4 \pi}
	\left[ \log(m_D R) + \gamma \right],
\label{f2nad}
\end{eqnarray}
where $\gamma$ is Euler's constant.  The terms proportional to $R$
can be absorbed into an order-$g^2$ correction to the Debye mass in
the $g^4$ term in (\ref{CPL}), which is proportional to $\exp(-2 m_D R)$.
The terms proportional to $R \log R$, however, cannot be absorbed
into the Debye mass.  Nadkarni concluded that the next-to-leading
order correction to the Debye mass is undefined.

We now show how the next-to-leading order Debye mass can be extracted from
the calculation of the correlator of Polyakov loops.
The $R \log R$ terms in (\ref{f1nad})
and (\ref{f2nad}) indicate that the integrals  in (\ref{f1def})
and (\ref{f2def}) are sensitive to the
infrared region.  As a regularization of the infrared region,
we replace the massless propagator $1/{\bf p}^2$ in (\ref{f1def})
by $1/({\bf p}^2+ m_{\rm mag}^2)$, and similarly for
$1/({\bf p}_1 - {\bf p}_2)^2$ in (\ref{f2def}).
If we first set $m_{\rm mag} = 0$ and then evaluate the integrals in the limit
$R \to \infty$, we recover Nadkarni's results (\ref{f1nad}) and (\ref{f2nad}).
However if we take the limit $R \to \infty$ in the presence of the
regulator, we obtain a different result.

We first consider $f_1(R)$ in (\ref{f1def}).
Evaluating the first integral over ${\bf p}$ with dimensional regularization,
we obtain the constant value $- m_D/4 \pi$.
The second integral over ${\bf p}$ gives a function of ${\bf k}$
that has branch points at $k = \pm i (m_D + m_{\rm mag})$.
The integral over ${\bf k}$ can be evaluated by expressing it as an
integral over $k$ along the entire real axis, and then deforming the contour
into the upper-half complex plane.  It receives
contributions from the double pole at $k = i m_D$ and from the branch cut
beginning at $k = i(m_D + m_{\rm mag})$.  In the limit $R \to \infty$,
the contribution from the branch cut is exponentially suppressed by
a factor of $e^{- m_{\rm mag}R}$, and can therefore be neglected.
The second integral over ${\bf p}$ in (\ref{f1def}) can therefore
be replaced by its value at $k = i m_D$, and it reduces to
$\log(2 m_D/m_{\rm mag})/(8 \pi m_D)$ in the limit $m_{\rm mag} \ll m_D$.
Evaluating the remaining contour
integral over $k$, we find that the asymptotic behavior of $f_1(R)$
as $R \to \infty$ is
\begin{equation}
f_1(R) \; \longrightarrow \; {m_D R \over 4 \pi}
	\left[ - 2 \log{2 m_D \over m_{\rm mag}} + 1\right],
\label{f2reg}
\end{equation}

We next consider $f_2(R)$ in (\ref{f2def}).  In the limit $m_{\rm mag} \to 0$,
the integral over ${\bf p}_1$ and ${\bf p}_2$ gives
$\log((k^2 + 4 m_D^2)/4 m_D^2)/[16 \pi^2 k^2 (k^2 + 4 m_D^2)]$,
which has coincident poles and logarithmic branch points at $k = \pm 2 i m_D$.
In the presence of the regulator, the integral has branch points at
$k = \pm 2 i m_D$ and $k = \pm i (2 m_D + m_{\rm mag})$.
The integral over ${\bf k}$ can again be evaluated by expressing it as an
integral over $k$ along a contour that is deformed into the
upper-half complex plane.  The contribution from the branch cut
beginning at $k = i (2 m_D + m_{\rm mag})$ is suppressed by a factor of
$e^{- m_{\rm mag} R}$ and can be neglected in the limit $R \to \infty$.
The contribution from the branch point beginning at $k = 2 i m_D$
has the asymptotic behavior
\begin{equation}
f_2(R) \; \longrightarrow \; {m_D^2 \over \pi m_{\rm mag}^2} \log R.
\end{equation}
There is a quadratic infrared divergence that is cut off by the
magnetic mass, but there are no terms linear in $R$.

Inserting (\ref{f2reg}) into (\ref{CPL}), we find that the correction
proportional to $R$ has precisely the form $-2 \delta m_D R$, where
$\delta m_D$ is given by Rebhan's expression (\ref{delmD}).
This correction can be absorbed into the order-$g^4$ term in (\ref{CPL})
by making the substitution $m_D \to m_D + \delta m_D$ in the potential $V(R)$.
It therefore represents a correction to the Debye mass.
This result was obtained by using a magnetic mass as an infrared regulator,
which separated the poles associated with the Debye mass
from the branch cuts associated with the emission of transverse gluons.
We could equally well have used a different infrared regulator,
such as dimensional regularization.  As long as the same regulator
is used in the calculation of the correlator of Polyakov loops
and in the calculation of the pole in the gluon propagator,
the next-to-leading order correction to the Debye mass will be the same.
This is evident from comparing the relevant parts of the integrals over
${\bf p}$ in (\ref{Pi-int}) and (\ref{f1def}).

The fact that the Debye mass depends on an infrared cutoff at
next-to-leading order should not be too disturbing.  At large $R$,
the correlator of Polyakov loops is dominated by a sum of
terms that fall exponentially like $\exp(-MR)$,
with $M$ being the mass of a color-singlet bound state
in (2+1)-dimensional extended QCD \cite{nadkarni2}.
This field theory is an $SU(3)$ gauge theory
describing magnetic gluons, coupled to an adjoint scalar field
describing electric gluons.  One of the bound states consists
predominantly of a pair of electric gluons, and its mass is
$2 m_D$ plus a small binding energy.  It is the contribution
of this bound state that corresponds to the perturbative contribution
to the Polyakov loop correlator calculated above.  The infrared
divergence in the next-to-leading order correction to $2m_D$
is a signal that the binding energy of this state
cannot be calculated in perturbation theory.

As an aside, it should be pointed out the truly asymptotic
behavior of the Polyakov loop correlator as $R \to \infty$
is dominated not by the lowest bound state of electric gluons,
but by a magnetic glueball \cite{nadkarni}.  Its coupling to the Polyakov
loop is suppressed by extra powers of the coupling constant,
but its mass is much less than $2 m_D$ at weak coupling,
so it dominates at sufficiently large $R$.

We have calculated the Debye mass for a quark-gluon plasma to
next-to-leading order in the QCD coupling constant from the
gauge-invariant correlation function of Polyakov loops.
The result agrees with that obtained by Rebhan from the pole
in the gluon propagator.  It is gauge invariant, but depends
logarithmically on nonperturbative effects associated
with the screening of static chromomagnetic fields.
This result provides support for the general definition of the
Debye mass in terms of the location of the pole in the propagator of the
gauge field.

This work was supported in part by the U.S. Department of Energy,
Division of High Energy Physics, under Grant DE-FG02-91-ER40684.

\vfill\eject

\end{document}